# High-pressure/high-temperature phase diagram of zinc

D. Errandonea[1], S. G. MacLeod[2,3], J. Ruiz-Fuertes[4], L. Burakovsky[5], M. I. McMahon[3]

C. W. Wilson[2], J. Ibañez[6], D. Daisenberger[7], and C. Popescu[8]

[1]*Departamento de Física Aplicada-ICMUV, Universidad de Valencia, MALTA Consolider Team, Edificio de Investigación, C/Dr. Moliner 50, 46100 Burjassot, Valencia, Spain*

[2]*Atomic Weapons Establishment, Aldermaston, Reading, RG7 4PR, United Kingdom*

[3]*SUPA, School of Physics and Astronomy, and Centre for Science at Extreme Conditions, The University of Edinburgh, Edinburgh, EH9 3FD, United Kingdom*

[4]*DCITIMAC, MALTA Consolider Team, Universidad de Cantabria, 39005 Santander, Spain*

[5]*Theoretical and Computational Physics Divisions, Los Alamos National Laboratory, Los Alamos, New Mexico 87545, USA*

[6]*Institute of Earth Sciences Jaume Almera, CSIC, 08028 Barcelona, Spain*

[7]*Diamond Light Source Ltd., Harwell Science and Innovation Campus, Didcot, Oxfordshire OX11 0DE, United Kingdom*

[8]*CELLS-ALBA Synchrotron Light Facility, Cerdanyola, 08290 Barcelona, Spain*

**Abstract:** The phase diagram of zinc (Zn) has been explored up to 140 GPa and 6000 K, by combining optical observations, x-ray diffraction, and *ab-initio* calculations. In the pressure range covered by this study, Zn is found to retain a hexagonal close-packed (hcp) crystal symmetry up to the melting temperature. The known decrease of the axial ratio (*c/a*) of the hcp phase of Zn under compression is observed in x-ray diffraction experiments from 300 K up to the melting temperature. The pressure at which *c/a* reaches $\sqrt{3}$ (≈10 GPa) is slightly affected by temperature. When this axial ratio is reached, we observed that single crystals of Zn, formed at high temperature, break into multiple polycrystals. In addition, a noticeable change in the pressure dependence of *c/a* takes place at the same pressure. Both phenomena could be caused by an isomorphic second-order phase transition induced by pressure in Zn. The reported melt curve extends previous results from 24 to 135 GPa. The pressure dependence obtained for the melting temperature is accurately described up to 135 GPa by using a Simon–Glatzel equation: $T_m = 690\,K\,(1 + \frac{P}{10.5\,GPa})^{0.76}$, where *P* is the pressure in GPa. The determined melt curve agrees with previous low-pressure studies and with shock-wave experiments, with a melting temperature of 5060(30) K at 135 GPa. Finally, a thermal equation of state is reported, which at room-temperature agrees with the literature.

## 1. Introduction

The behavior of structural and mechanical properties of metals under high-pressure (HP) and high-temperature (HT) conditions has received considerable attention for decades. In particular, the knowledge of this behavior has many implications for materials science and Earth and planetary sciences. Great efforts have been devoted during the present century to the study of iron [1 - 4], titanium [5, 6], molybdenum [7 – 10], copper [11 - 13], magnesium [14, 16], tantalum and tungsten [16 – 18] among other metals. In contrast, much less attention has been paid to zinc (Zn) and the elements in group 12 of the periodic table. Zn is a metal that crystallizes in the hexagonal close-packed (hcp) structure (space group $P6_3/mmc$). However, the hcp structure of Zn has an unusually large axial ratio ($c/a$) when compared with most hcp metals [19]. Zinc is the 24$^{th}$ most abundant element in the Earth's crust and has multiple applications in the automobile, construction, and ship-building industries among others [20].

The compressibility of Zn at room-temperature (RT) was first studied by Bridgman in 1941 using a piston-cylinder apparatus [21]. His results were confirmed by Vaidya and Kennedy in 1970 [22]. An equation of state (EOS) up to 1.5 Mbar (150 GPa) was determined by McQueen and Marsh from shock-wave experiments in 1960 [23]. Additionally, the melt curve was determined up to 6 GPa from differential thermal analysis measurements in 1973 [24]. After these pioneering works, a resurgence of interest in the behavior of Zn under HP conditions took place in 1995. This was triggered by a Mössbauer spectroscopy study which suggested that an electronic topological transition occurs in Zn at 6.6 GPa [25]. This work was followed by a series of studies using different techniques [19, 26 – 34]. All these studies were carried out at RT (or below it) and were focused on compressibility, the supposed anomaly of $c/a$, and the proposed electronic topological transition. In contrast with RT studies, the continuation of HP-HT studies had to wait until the last decade. Melting studies have been carried out up to 24 GPa [35, 36] and very recently resistivity measurements were conducted up to 5 GPa and 900 K [37] and measurements of the structure of liquid Zn were performed up to 1.6 GPa and 800 K [38]. A second shock-wave experiment on Zn has also been reported in 2012 [39]. In contrast with experiments, theoretical efforts have been focused only on 0 K calculations [40 - 44].

In order to deepen knowledge of the behavior of Zn under HP-HT conditions we report here a combined experimental and theoretical study. Melting measurements have been performed up to 80 GPa and 3800 K, x-ray diffraction (XRD) studies up to 16 GPa and 1000 K, and *ab-initio* calculations up to 140 GPa and 6000 K. The melt curve of Zn is accurately determined, extending it from 24 GPa to 135 GPa. A pressure-volume-temperature (P-V-T) EOS is determined, and the changes induced by pressure and temperature in the crystal structure are discussed. The results are compared with previous studies summarized in this introduction. The reliability of the present study is supported by the consistency between experiments and calculations.

## 2. Details of experiments and calculations

The melt curve of Zn was measured in a laser-heated diamond-anvil cell (DAC) [44] using the laser-speckle method and the temperature versus laser-power method [35, 44]. The experiments were performed in a Boehler-Almax DAC whose anvils have diamond culets of 250 μm in diameter. In order to load the sample, we used a tungsten gasket pre-indented to 40 μm with a hole of 80 μm diameter in the center. As a sample, we used a fresh piece of Zn obtained from a 10-μm thick foil (Alfa Aesar, 99.9% purity). The sample was embedded in MgO, which acts as a thermal insulator from the diamond anvils and also as pressure-transmitting medium. For the pressure range of our experiments, MgO has a higher melting temperature than the temperature coverage of our experiments [45]. In addition, there are no reports in the literature of chemical reactions between MgO and metals under HP-HT conditions up to 5000 K [7, 45]. In order to guarantee the chemical integrity of the sample, measurements at selected pressures were carried out on different samples. In addition, heating was constrained to a time period smaller than 15 s. The sample was heated from one side using a laser-heating setup equipped with a Nd:YLF laser. A series of five experiments were performed at each reported pressure on different portions of the samples. The reported melting temperature ($T_m$) is the average value and the error is assumed to be the absolute value of the maximum deviation from the average. Temperatures were measured by spectroradiometry, fitting the thermal emission of the sample to the Planck equation and assuming a grey-body radiation with wavelength independent emissivity [46, 47]. Pressure was determined at RT using the ruby scale [48]. Pressures measured before and after heating agree within 0.5 GPa. The thermal pressure was neglected since it is theoretically estimated to be smaller than 1 GPa [49].

For the XRD studies we used resistively-heated membrane-type DACs equipped with diamonds with 300 μm culets and loaded with NaCl, which acted as the pressure-transmitting medium and pressure standard [50]. Rhenium gaskets were used to contain the sample and pressure medium. The Zn sample was obtained from the same foil used for the laser-heating experiments. The set-up used was similar to the one previously described by Stinton *et al.* [14] and Cazorla *et al.* [51]. Powder angle-dispersive XRD measurements were carried out following several P-T paths (described in the next section). These experiments were conducted using the MSPD-BL04 beamline at the ALBA synchrotron [52] and the I15 beamline at Diamond Light Source (DLS). We used an x-ray wavelength of 0.4246 Å and a beam size of 15 μm × 15 μm at ALBA, and 0.4246 Å and a beam size of 30 μm (round diameter) at DLS, respectively. In addition to the HT measurements one experiment at RT was performed to compare with previous studies. The temperature was measured using a K-type thermocouple attached to one of the diamond anvils, close to the gasket. The accuracy of the temperature measurements is better than 0.4%. The accuracy in pressure determination was 0.1 GPa. The 2D XRD patterns were collected on a Rayonix charge-coupled device (CCD) detector at ALBA and a MAR345 image plate detector at DLS. A rocking (±3°) of the DAC was used to improve the homogeneity of the Debye rings. The exposure time was 10 s for every XRD pattern. The 2D images of the CCD were integrated with the FIT2D software [53]. The resulting 1D diffraction patterns were analyzed by means of a Le Bail fitting [54] using FullProf [55].

The structural stability and melting of Zn was also studied using density-functional theory (DFT) calculations, which were implemented in the Vienna *Ab initio* Simulation Package (VASP). For the melt curve, the Z method was employed [56 - 58]. The quantum molecular dynamics runs took 15 - 25 ps (15000 – 25000 time steps). We used the generalized-gradient approximation (GGA) with the Perdew-Burke-Ernzerhof (PBE) exchange-correlation functional. In order to model Zn, we employed the $[Ar]3d^{10}4s^2$ electron core-valence representation. The valence electrons were described with a plane-wave basis set with a cutoff energy of 440 eV, while the core electrons were represented by projector augmented-wave (PAW) pseudopotentials. For different volumes, corresponding to pressures from 0 to 140 GPa, an optimization of the unit-cell parameters, *c* and *a*, at a constant volume was carried out. At 0 GPa the calculated *c/a* ratio (1.860) agrees within 0.2% with the value reported by Kenichi *et al.* [19, 29], which also agrees with the value obtained from our experiments (*c/a* = 1.856). Systems of 512 atoms (8 x 8 x 4) of different unit-cell volumes were used for the melting simulations, with a single Γ-point. For such large systems, convergence to 2.5 meV/atom was achieved in the whole interval of pressures, which corresponds to ~ 30 K uncertainty in the value of the corresponding melting temperature. Such a small uncertainty is well within the error bars of the Z method itself.

3. **Results and discussion**

    a. **Melt curve**

In Fig. 1 we show (as solid triangles) the melting temperatures determined at different pressures from our laser-heated DAC experiments. At low-pressure (P < 24 GPa) the present results agree with those previously reported in the literature [24, 35, 36]. They also agree well with an extrapolation of the Simon equation fitted by Errandonea [36] from previous published data (black dotted line in Fig.1). In the figure we also represent the melt curve obtained from DFT calculations up to 135 GPa (red squares). The calculated melt curve agrees with the experiments up to the highest measured pressure (80 GPa). There is also good agreement with shock-wave experiments at 120 GPa [39]. The consistency between calculations and different experiments improves confidence in the reliability of the reported melt curve. A fitting using the Simon equation [59] to results from present and previous static-compression experiments, DFT calculations, and shock-wave experiments leads to the following melting relation $T_m = 690\ K\ (1 + \frac{P}{10.5\ GPa})^{0.76}$, where *P* is the pressure in GPa. This Simon fit is shown as a solid red line in Fig. 1. The melt curve of Zn initially rises steeply as a function of pressure, though at pressures above 60 GPa the slope (*dT$_m$/dP*) decreases slightly . As a consequence of this, the Simon function used previously to describe the melt curve of Zn [36] (fitted from data collected below 24 GPa) tends to overestimate the melting temperature above 80 GPa. However, the present fit properly describes all the data available on melting of Zn.

From our calculations we also determined the Hugoniot curve [60] of Zn. The results are shown in Fig. 1 (red dot-dashed line). The calculated Hugoniot follows a similar dependence to the experimental Hugoniot reported by McQueen and Marsh [23] more than half a century ago. However, these experiments show a minor kink in the Hugoniot near 50 GPa and 1375 K. The presence of such a discontinuity is not reproduced in our calculations. A kink in the Hugoniot below the melting temperature can only be caused by a phase transition to a further solid phase. Neither our calculations, nor our experiments found evidence to confirm the existence of a solid-solid phase transition in Zn induced by pressure at HT at pressures close to 50 GPa. In particular, the existence of a solid-solid phase transition is contradicted by the smooth behavior here determined for the melting curve. On the other hand, structural changes associated to solid-solid transitions can be detected by visual observation using the laser speckle technique here employed to determine the melt temperature [61, 62]. We have performed one of these experiments at 50 GPa and five at higher pressures and none of them provided evidence of a solid-solid phase transition up to the melt temperature. These results highlight the necessity for new shock-wave measurements in Zn using laser-driven shock techniques [63] to confirm that beyond 50 GPa Zn remains hexagonal at HT up to the melting temperature.

b. **X-ray diffraction**

We performed XRD studies under compression up to 16 GPa at RT and also at HT. The intention of the studies was two-fold: to explore the influence of temperature on the behavior of the *c/a* ratio under compression and to determine a P-V-T EOS for Zn. At RT we found results which agree fully with previous studies [26, 27, 32]. In our case *c/a* continuously decreases under compression, reaching a value of $\sqrt{3}$ around 10 GPa. In particular, the two linear compressibilities determined at RT and ambient pressure are $k_a = 2.3 \; 10^{-2}$ GPa$^{-1}$ and $k_c = 9.8 \; 10^{-2}$ GPa$^{-1}$. Qualitatively similar results were found when different isotherms were collected. In Fig. 2 we show a selection of XRD patterns measured at different pressures, at temperatures between 455 and 470 K. In the figure, the relative difference between the evolution of the (002) and (100) Bragg reflections can be observed, which is a consequence of the change of *c/a* under compression. At 5.05 GPa the (002) reflection is at a lower angle than the (100) reflection; because $c/a > \sqrt{3}$. Under compression the (002) peak moves faster than the (100) peak due to the fact that *c* is considerably more compressible than *a*. As a consequence, both peaks overlap at 9.25 GPa, and at 12.5 GPa the (002) peak is at a higher angle than the (100) peak; because $c/a < \sqrt{3}$. From the experiments, we determined the pressure dependence of *c/a*.

In Fig 3 we plot $1 - \sqrt{3}a/c$ versus pressure. The results correspond to experiments carried out at 300 and 600 K. This figure clearly shows the rapid change of *c/a* and the fact that $c/a = \sqrt{3}$ (which corresponds to $1 - \sqrt{3}a/c = 0$) is reached at a "critical" pressure ($P_C$) of 10 GPa at 300 K and at 10.4 GPa at 600 K. We found that when isotherms at 460, 740, and 850 K are followed, $c/a = \sqrt{3}$ is also reached near 10 GPa, with a slight tendency to increase $P_C$ as temperature is raised. From calculations we got a similar behavior for the *c/a* ratio than from experiments. Results from calculations at T = 0 K are included in Fig. 3 to demonstrate that calculations reproduce the kink in *c/a* near 10 GPa. After $c/a = \sqrt{3}$ is reached we found that the pressure dependence of *c/a* changes because of a sudden reduction of the linear compressibility of the *c*-axis. This is clearly illustrated by the inset of Fig. 3 where the pressure derivative of *c/a* is shown for the experiment carried out at 300 K. In the inset it can be appreciated the abrupt change of the pressure dependence of *c/a*. The existence of this kink (slope change) in the pressure dependence of *c/a* has been attributed to a softening of acoustic modes [35] and can be caused by an isomorphic phase transition [64]. The fact that there is no detectable discontinuity in either the unit-cell volume or the *c/a* ratio suggest that the proposed phase transformation might be a second-order transition [65, 66]. To conclude with the discussion of the behavior of the *c/a* axial ratio, we would like to mention that calculations found that the ideal ratio $c/a = \sqrt{8/3}$ is reached at 50 GPa, which is in good agreement with previous experiments [32].

Coincident with the change in the behavior of *c/a*, we noticed another interesting observation. At HT the polycrystalline Zn foil recrystallized into a single crystal. This is shown in Fig. 4, where the (002) and (100) single crystal reflections are shown in a CCD image measured at 4.8 GPa and 600 K. The same kind of single crystal diffraction pattern is found at 7.2 GPa and 600 K. However, when the pressure exceeds 10 GPa, the Zn crystal breaks into multiple crystallites, and the appearance of the corresponding diffraction patterns transform from single-crystal spots to highly textured Debye rings. This is illustrated in the CCD images collected at 10.4 and 13.2 GPa and 600 K. This phenomenon is fully consistent with the occurrence of a phase transition proposed based upon the behavior of *c/a*. Such transition would take place between isomorphic hexagonal structures, which we will call hcp (P < 10 GPa) and hcp' (P > 10 GPa). The first structure (hcp) possesses a ratio $c/a \geq \sqrt{3}$ with a larger axial compressibility along *c* than along *a*. The second structure (hcp') possesses a ratio $c/a \leq \sqrt{3}$ with a similar compressibility along *c* and *a*. The proposed second-order transition is consistent with phonon softening [65] proposed by calculations [35] and with changes reported for the resistivity of Zn by Lynch and Drickamer [67].

In Fig. 5 we summarize the findings of the XRD experiments. The black squares represent all the P-T conditions at which the hcp phase was detected and the white squares the P-T conditions at which the hcp' phase was detected. According to the experiments, the phase boundary (dashed line) is nearly vertical and has a positive slope. Based upon the Clausius–Clapeyron equation [68], the steep phase boundary implies that the specific entropy change of the phase transition is nearly zero. This observation is consistent with the occurrence of a second-order transition. Regarding the positive slope ($dT/dP$) of the hcp-hcp' boundary, we would like to comment that solid-solid transitions could possess either a positive or a negative slope [69]. Examples of both cases are the γ–ε and α–ε transitions of iron [70]. Following Clausius–Clapeyron, the experimentally determined positive slope suggests that both the volume and the entropy decrease at the transition.

In Fig. 5, we represent with open diamond P-T points where a rapid recrystallization of Zn was detected. Such a phenomenon is commonly found in the pre-melting state [1]. The melting is detected at a slightly higher temperature than the disappearance of the Bragg reflections of Zn. The P-T points corresponding to the detection of melting by XRD are shown as solid black diamonds in Fig. 5. The melting temperature detected by XRD is in good agreement with previous studies [24, 35, 36]. For consistency, melting points detected in the DAC experiments at P < 16 GPa are also included in Fig. 5 (solid triangles). In order to illustrate the detection of melting in the XRD experiments, Fig. 6 shows a selection of XRD patterns collected in a heating sequence. The conditions gradually changed from 6.3 GPa and 300 K to 5.7 GPa and 953 K. At these conditions the melting of Zn is detected by means of the disappearance of the Bragg reflections of Zn and the appearance of weak diffuse scattering.

c. **P-V-T EOS**

From the XRD experiments we determined the unit-cell parameters for Zn at different pressure and temperature conditions. The results obtained from five isotherms are shown in Fig. 7. From the P-V data measured at RT we performed a third-order Birch-Murnaghan EOS fit to the data [71]. Since there is no discernible atomic volume discontinuity between hcp and hcp' the data collected in the pressure range have been included in the fit, as previously reported by other authors [27]. From the fit we obtained the ambient pressure bulk modulus $B_0$ = 63(2) GPa, and its pressure derivative $B_0$' = 5.6(4) as well as the unit-cell volume $V_0$ = 30.6(1) Å$^3$. These parameters are in good agreement with those previously reported [27]. From an isobar measured at 6.2(2) GPa, we have determined the linear thermal expansion of Zn, obtaining $\alpha_a$ = 15 10$^{-6}$ K$^{-1}$ and $\alpha_c$ = 65 10$^{-6}$ K$^{-1}$. These values are very similar to those determined at ambient pressure [72], which suggests that the thermal expansion of the hcp phase is barely affected by pressure. In addition, our results indicate that the thermal expansion along the *c*-axis is approximately four times larger than the thermal expansion along the *a*-axis. Noticeably, as discussed above, $k_c/k_a$ is also approximately four. This result agrees with the understanding that the thermal expansion and compressibility are closely correlated with the nature of the chemical bonding and the crystallographic structure of materials [73].

    Finally, we found that the isotherms shown in Fig. 7 can be described using the Birch–Murnaghan isothermal formalism [74]. The calculated isotherms are shown in the figure together with the experimental results. In the fit of the P-V-T EOS we assumed that the volumetric thermal expansion and pressure derivative of the bulk modulus are temperature independent and that only the bulk modulus is temperature dependent having a linear dependence: $B_0(T) = B_0(300\,K) + \beta\,(T - 300\,K)$, with $\beta$ the only fitting parameter. For the volumetric thermal expansion we used the value determined from the linear expansion coefficients assuming $\alpha = 2\alpha_a + \alpha_c = 95$ 10$^{-6}$ K$^{-1}$. For $V_0$, $B_0$, and $B_0$' at 300 K we used our own values: 30.6 Å$^3$, 63 GPa, and 5.6, respectively. As can be seen in Fig. 7 the assumed approximations are sufficient for describing the evolution of volume with P and T in the P-T range covered by our studies. The determined value for $\beta$ was –

9.8(9) $10^{-3}$ GPa/K, which indicates a decrease of the bulk modulus as the temperature increases. The results of the P-V-T EOS fit are summarized in Table I.

## 4. Concluding Remarks

We have performed a combined experimental and theoretical study of Zn under HP-HT conditions. Significantly, we have extended the melt curve of Zn from 24 GPa to 140 GPa. The melt curve reported here includes all previous measurements, in particular those obtained from shock-wave experiments. We have also studied the influence of pressure and temperature on the crystal structure of Zn. We have shown that the known anisotropic compressibility of Zn at RT is repeated at high temperature. We also found a change in the behavior of the axial ratio of hcp Zn at 10 GPa, which is the critical pressure at which *c/a* reaches the value $\sqrt{3}$. Results from five different isotherms indicate that this critical pressure is independent of temperature. This observation occurs together with a drastic change of the XRD patterns of Zn; from a single-crystal pattern to a powder XRD pattern. A possible explanation for both phenomena is the existence of a second-order isostructural phase transition. No evidence of other solid-solid transitions was found up to 140 GPa. A RT EOS for Zn has been determined and is in agreement with previous studies [29]. The thermal expansion has also been determined and a P-V-T EOS was used to describe the pressure dependence of the volume at different isotherms. It has been found that the bulk modulus of Zn decreases with temperature.

**Acknowledgments**

The authors are thankful for the financial support to this research from the Spanish Ministerio de Economía y Competitividad, the Spanish Research Agency, and the European Fund for Regional Development under Grant Nos. MAT2016-75586-C4-1-P,

**Table I:** Results of fitting the experimental volumes with the P-V-T EOS model described in the text.

| $V_0$ (Å$^3$) | $B_0$ (GPa) | $B_0'$ (dimensionless) | $\alpha$ (K$^{-1}$) | $\beta$ (GPa/K) |
|---|---|---|---|---|
| 30.6(1) | 63(2) | 5.6(4) | 95 10$^{-6}$ | 9.8(9) 10$^{-3}$ |

**Figure 1:** Melt curve of Zn. Triangles are experimental data points from the present study. The red squares represent the calculated melting temperatures at different pressures. The red solid line is the present Simon fit. White, gray, and black circles are melting points from previous studies [24, 35, 36]. The black dotted line is the Simon fit previously reported by Errandonea [36]. The black diamond corresponds to melting determined in shock-wave experiments [39]. The Hugoniot measured by McQueen is represented by open diamonds [23] and the calculated Hugoniot is represented by a dot-dashed red line.

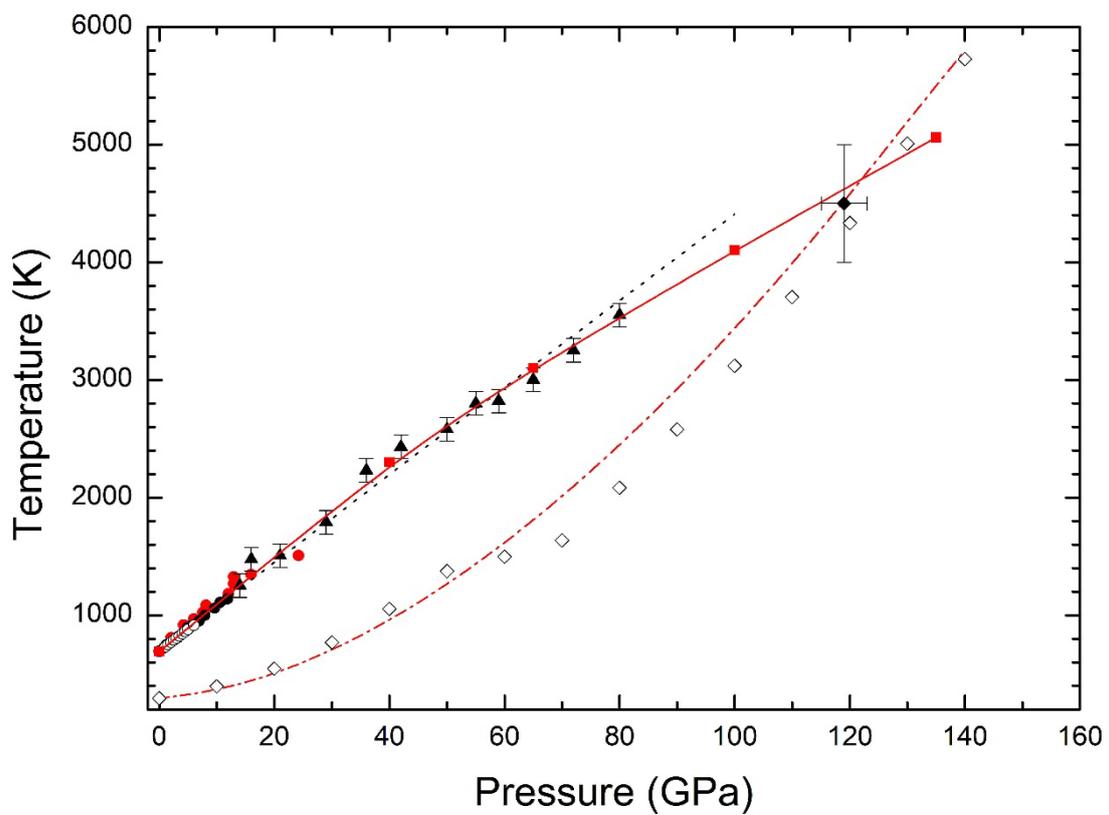

**Figure 2:** X-ray diffraction patterns collected at different pressures following a 460(10) K isotherm. Pressures and temperatures are indicated. In the lowest trace the NaCl and Zn Bragg reflections are identified. The three main peaks of Zn are labeled in all patterns to show the different pressure evolution due to the anisotropic compression.

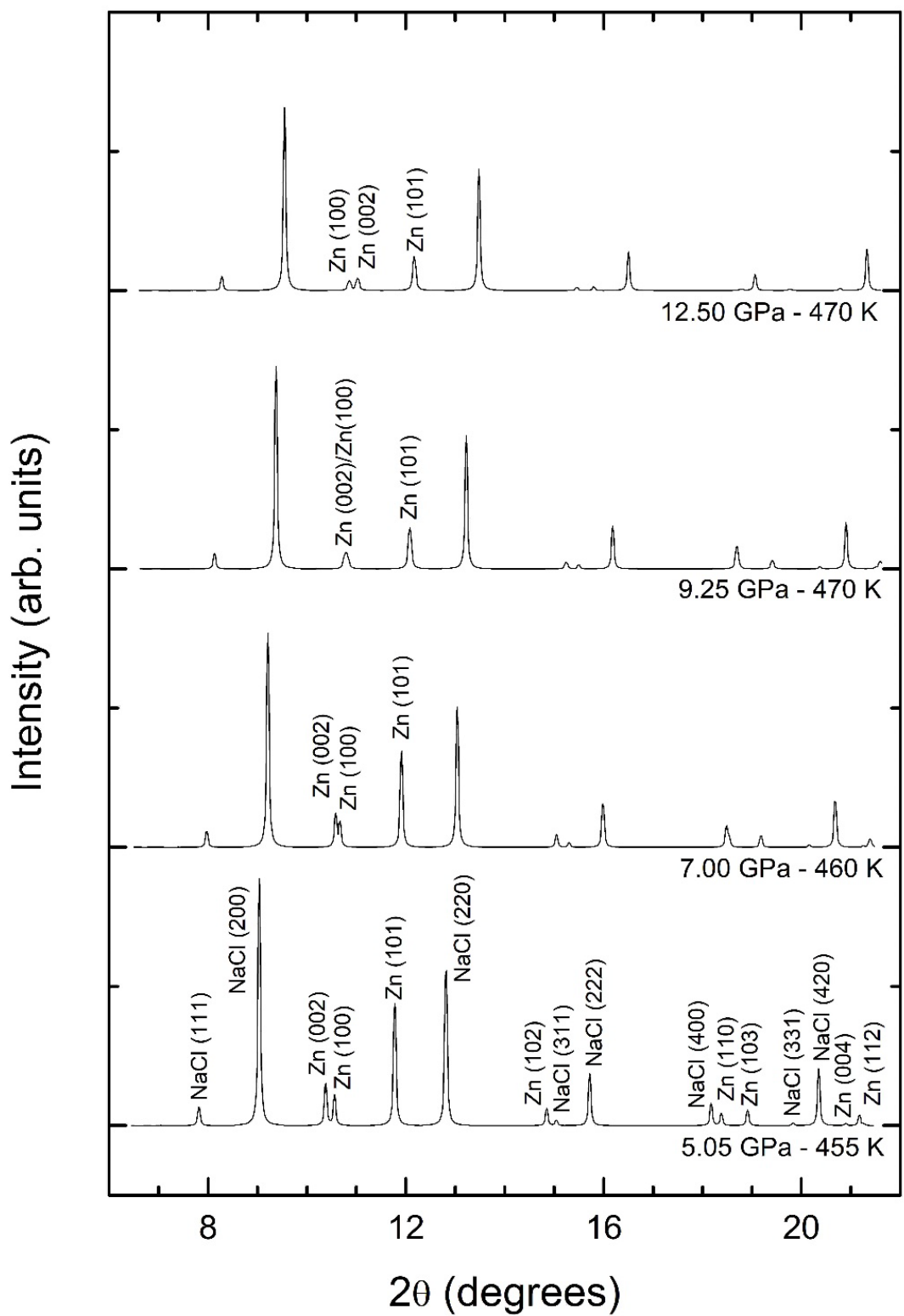

**Figure 3:** $1 - \sqrt{3}a/c$ versus pressure. The results represented by squares are obtained from experiments carried out at 300 and 600 K. The value zero ($\sqrt{3} = c/a$) is reached at a "critical" pressure ($P_C$) of 10 and 10.4 GPa, respectively. At this pressure a kink in the curve can be observed. Solid lines are polynomial fits to the experimental results for $P > P_C$ and $P < P_C$. Similar results are obtained at other temperatures. Results from calculations at T = 0 K are shown by open circles.. The inset show the pressure derivative of *c/a* obtained from a fit to the 300 K experimental results. The discontinuity in the slope of *c/a* versus pressure can be clearly seen.

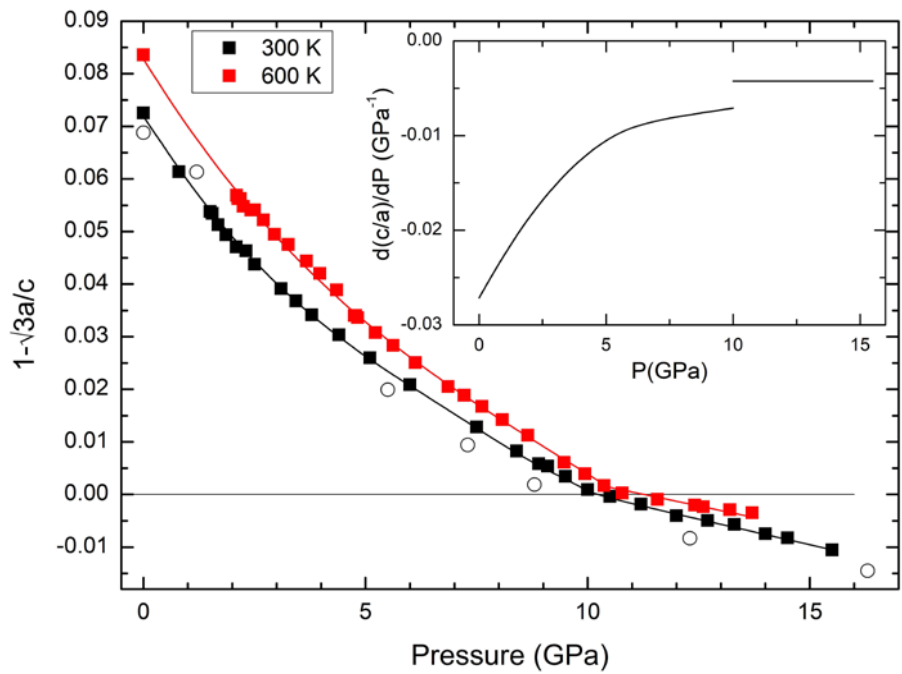

**Figure 4:** Images of parts of the CCD detector collected at different pressures following the 600(10) K isotherm. At 4.8 and 7.2 GPa Bragg reflections from a single crystal are observed. At 10.4 and 13.3 GPa powder-like rings are observed.

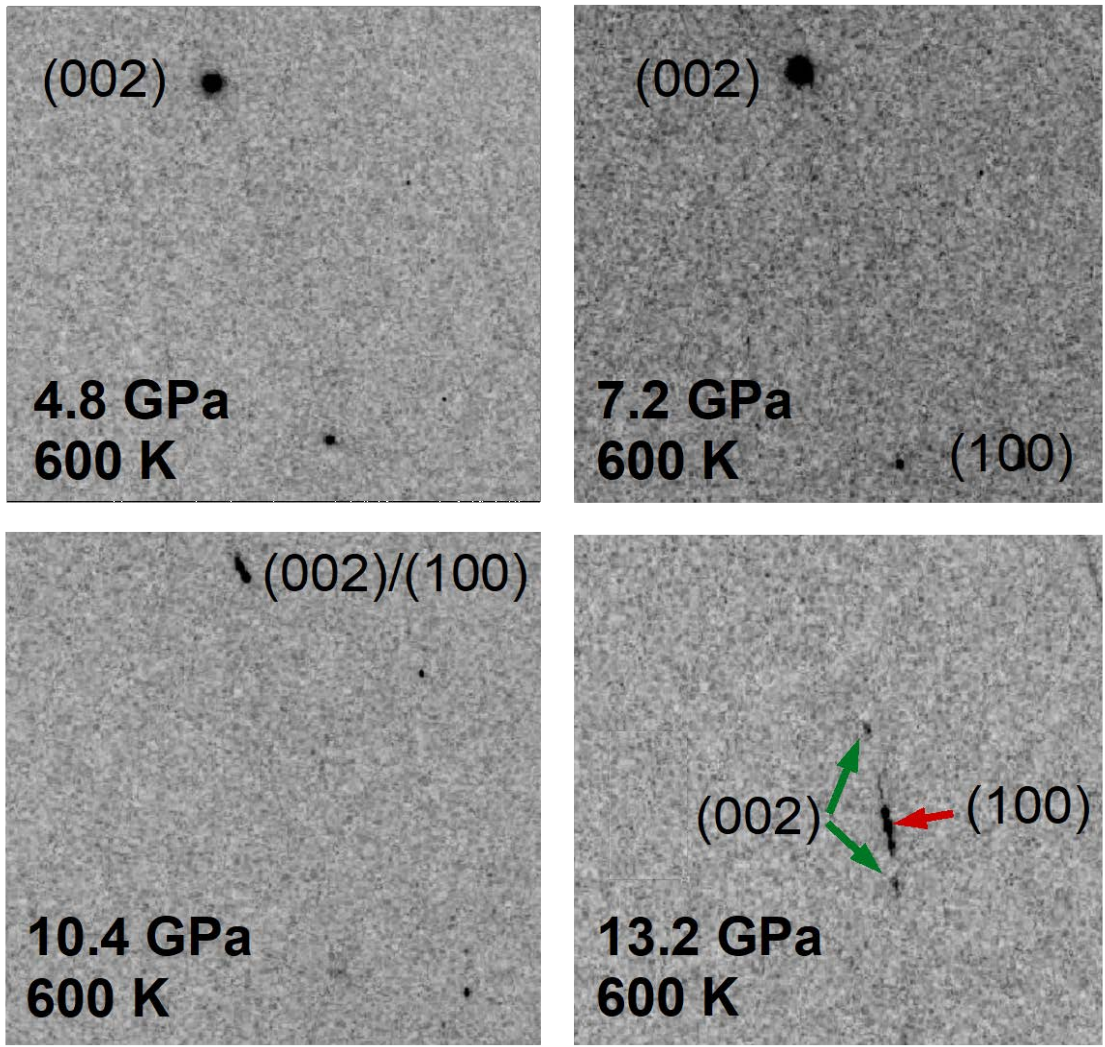

**Figure 5:** P-T phase diagram for P < 16 GPa and T < 1600 K. Square data points represent results from the XRD experiments. Solid squares are used for the hcp structure when $c/a > \sqrt{3}$ (hcp) and empty symbols when $c/a < \sqrt{3}$ (hcp´). White, red, and black circles are melting points from previous studies [24, 35, 36]. The triangles are melting points determined in the present laser-heating measurements. Empty diamonds correspond to rapid recrystallization and solid diamond to melting determined from XRD experiments. The solid line is the Simon fit to the melt curve and the dashed line the hcp-hcp´ phase boundary.

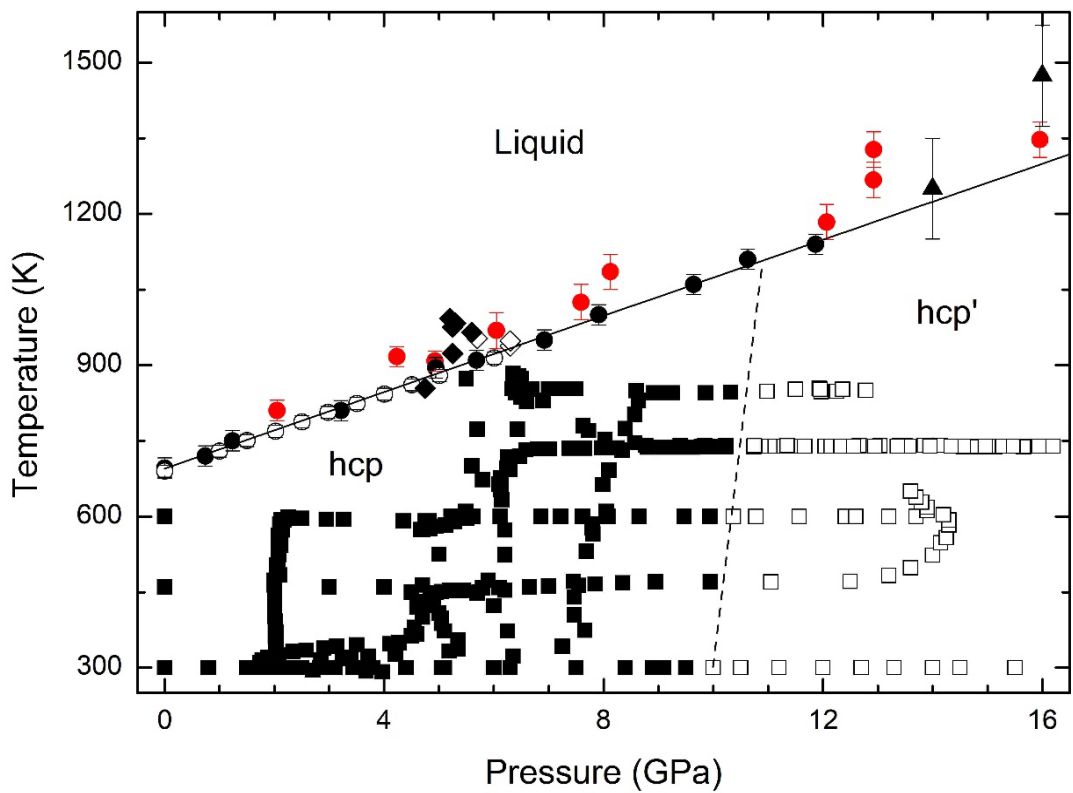

**Figure 6:** X-ray diffraction patterns measured at pressures near 6 GPa and with increasing temperature. Pressures and temperatures are indicated. In the lowest trace the NaCl and Zn Bragg reflections are identified. At 5.7 GPa and 953 K the peaks of Zn are lost due to melting. The diffuse scattering of the liquid can be seen at these conditions near 2θ = 12º.

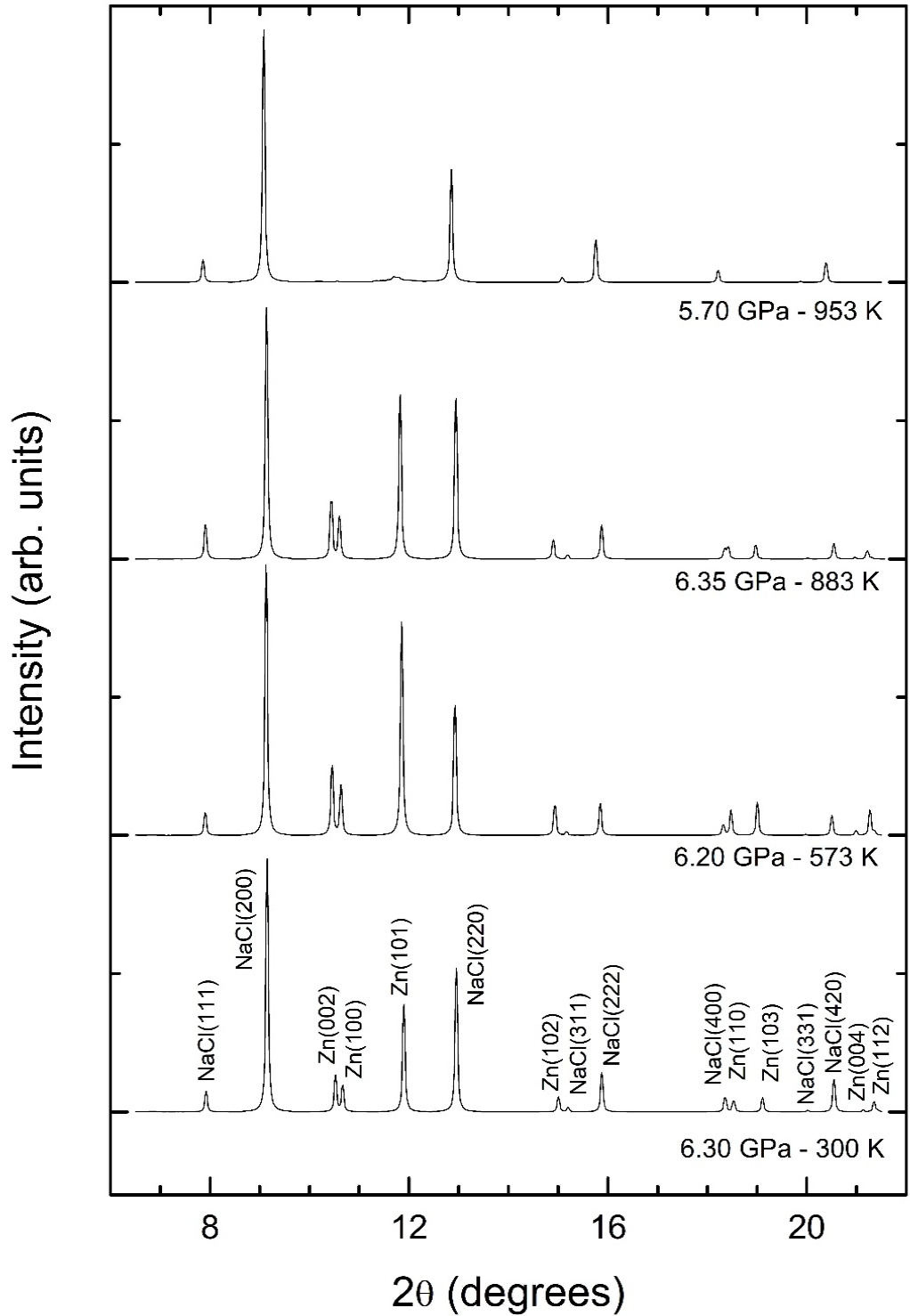

**Figure 7:** P-V-T EOS of Zn. Data corresponding to five isotherms are shown. The symbols represent experimental data points and the lines the calculated isotherms. At 740 and 850 K the isotherms are shown for pressures corresponding to solid Zn, and at lower pressures liquid Zn is observed.

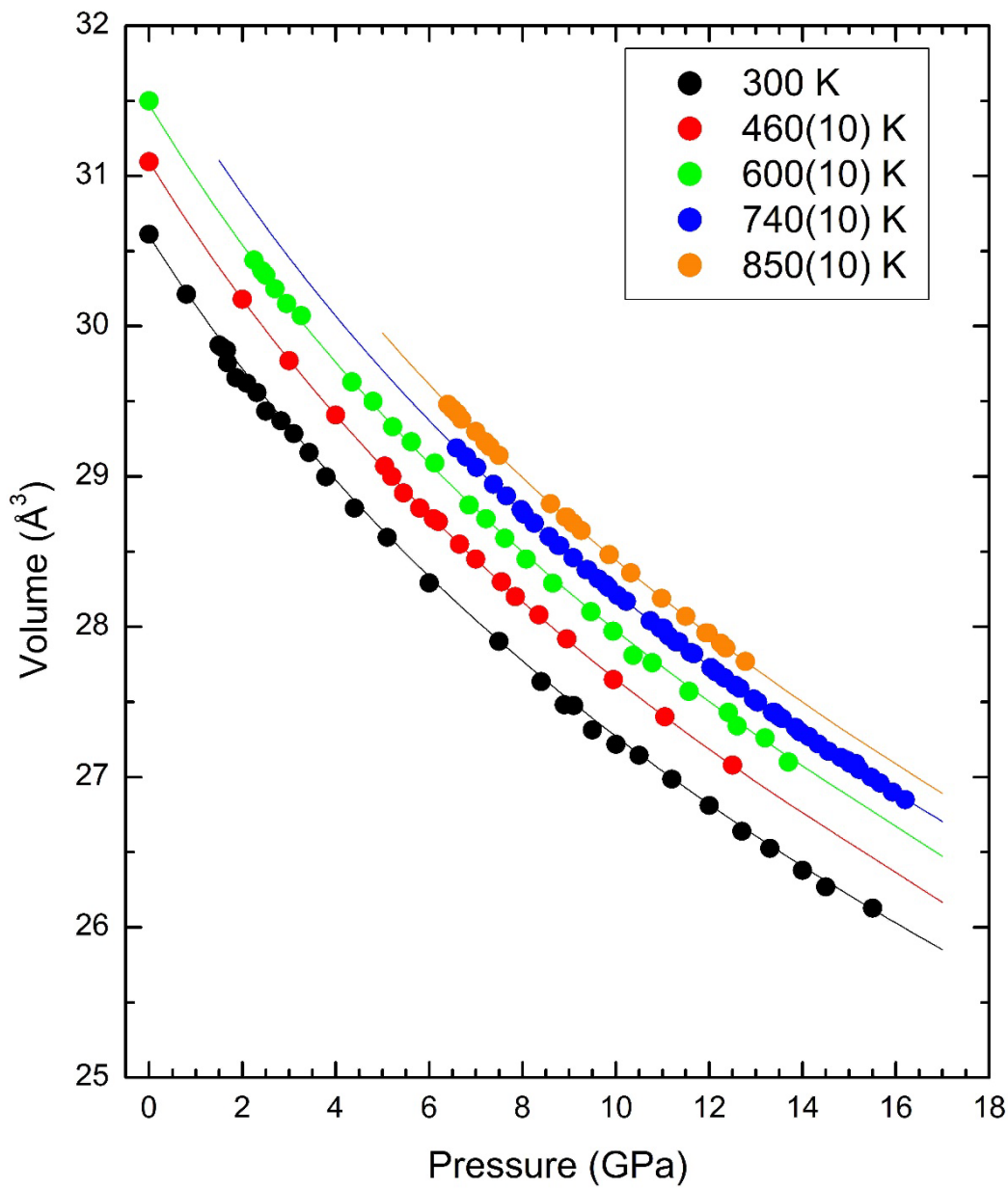